\documentclass[%
 reprint,
superscriptaddress,
 amsmath,amssymb,
prx,
]{revtex4-1}
\usepackage{booktabs}
\usepackage{multirow}
\usepackage{mathrsfs}
\usepackage{amsmath}
\usepackage{subfigure}
\usepackage{graphicx}
\usepackage{dcolumn}
\usepackage{bm}

\begin{document}

\preprint{APS/123-QED}

\title{Reference-Frame-Independent Quantum Key Distribution Using Fewer States}

\author{Hongwei Liu}
 \affiliation{School of Science and State Key Laboratory of Information Photonics and Optical Communications, Beijing University of Posts and Telecommunications, Beijing 100876, China}
 \affiliation{College of Liberal Arts and Science, National University of Defense Technology, Hunan, Changsha 410073, China}
\author{Jipeng Wang}%
 \affiliation{School of Science and State Key Laboratory of Information Photonics and Optical Communications, Beijing University of Posts and Telecommunications, Beijing 100876, China}
 \affiliation{College of Liberal Arts and Science, National University of Defense Technology, Hunan, Changsha 410073, China}
\author{Haiqiang Ma}
 \email{hqma@bupt.edu.cn}
 \affiliation{School of Science and State Key Laboratory of Information Photonics and Optical Communications, Beijing University of Posts and Telecommunications, Beijing 100876, China}
\author{Shihai Sun}
 \email{shsun@nudt.edu.cn}
 \affiliation{College of Liberal Arts and Science, National University of Defense Technology, Hunan, Changsha 410073, China}

\date{\today}

\begin{abstract}
Reference-frame-independent quantum key distribution (RFI QKD) protocol can reduce the requirement on the alignment of reference frames in practical systems. However, comparing with the Bennett-Brassard (BB84) QKD protocol, the main drawback of RFI QKD is that Alice needs to prepare six encoding states in the three mutually unbiased bases ($X$, $Y$,and $Z$), and Bob also needs to measures the quantum state with such three bases. Here, we show that the RFI QKD protocol can be secured in the case where Alice sends fewer states. In particular, we find that transmitting three states (two eigenstates of the $Z$ basis and one of the eigenstates in the $X$ basis) is sufficient to obtain the comparable secret key rates and the covered distances, even when the security against coherent attacks with statistical fluctuations of finite-key size is considered. Finally, a proof-of-principle experiment based on time-bin encoding is demonstrated to show the feasibility of our scheme, and its merit to simplify the experimental setup.
\end{abstract}

\pacs{Valid PACS appear here}
\maketitle


\section{\label{sec:level1}Introduction}

Quantum key distribution (QKD) has been attracting major attentions due to its higher-level security for information privacy. Unlike the classical cryptography, QKD is based on the principles of quantum physics and one-time pad, which has been proved unconditional secure \cite{shannon1949one,Bennett1984,E91,LO91,SP00}. Secure communication via QKD is one of important applications of the quantum information science, which can be realized with current technologies. Nowadays, QKD has been studied not only in laboratories \cite{scarani2004quantum,Wang2005Beating,DECOY05,RRDPS,comandar2016quantum,Wang}, but also in companies \cite{commercial}. Some companies have started making hardware and doing field tests focused on it \cite{Sasaki:11,PhysRevX.6.011024}. 

For the practical applications, realistic security of QKD has been investigated to close the gap between the assumptions made in the security proofs and the actual implementations \cite{gisin2010proposal,braunstein2012side,lo2012measurement}. Experimentally, it has been implemented via optical means, achieving key rate of 13.72 megabits per second over 2-dB of standard optical fiber \cite{Yuan:18} and of 1.16 bits per hour over 404 km of ultralow-loss fiber in measurement-device-independent configuration \cite{yin2016measurement}. Most recently, a new scheme was proposed, which is a promising step towards overcoming the rate-distance limit of QKD without quantum repeaters and greatly extending the range of secure quantum communications \cite{lucamarini2018overcoming,PhysRevX.8.031043}. However, reducing the systems complexity is still a vital issue in real-life applications of QKD.

In most QKD systems, a shared reference frame is required between a sender (Alice) and a receiver (Bob): alignment of polarization states for polarization encoding, interferometric stability for phase encoding. Hence, an active reference frame calibration is needed to ensure the achievable secure key rate. Although additional alignment parts appear feasible, they will increase the complexity of practical systems,  and even lead to extra information leakage through these ancillary processes \cite{Jain2011Device}. Fortunately, a promising scheme, named reference-frame-independent (RFI) QKD, is proposed to eliminate the requirement of alignment \cite{Laing2010Reference,yin2014reference}. In this protocol, three orthogonal bases ($X$, $Y$, and $Z$) are required to encode the information, in which the $X$ and $Y$ are used as monitoring-bases to estimate eavesdropper (Eve)'s information, and $Z$ basis is generally used to generate the final key. The states in the $Z$ basis, such as the time-bin eigen-states, are naturally well-aligned, whereas the states in $X$ and $Y$ bases are allowed changing slowly in the quantum channel. Due to this significance, RFI QKD could be very useful in several scenarios, such as earth-to-satellite QKD \cite{1367-2630-15-7-073001} and path-encoded chip-to-chip QKD \cite{PhysRevLett.112.130501}. Several theoretical studies and experimental works have been reported \cite{liang2014proof,PhysRevA.94.062330,PhysRevLett.115.160502,Liu18,Wang:17}.

As mentioned above, the RFI QKD protocol needs six states in the processes of key distribution, whereas only four states are required in the Bennett-Brassard (BB84) QKD protocol. If is it possible that we use only four or even less states to complete the task and meanwhile keep the merit of RFI QKD protocol? A theoretical study have explored this possibility, whose advantage is that doing so may simplify the implementation, e.g., less randomness is required and possibly fewer optical elements are need. By exploiting the additional information gleaned from the mismatches basis statistics \cite{PhysRevA.90.052314,PhysRevA.92.032305,PhysRevA.93.042308}, Tamaki et al.~\cite{PhysRevA.90.052314} showed that three states (two eigenstates of $Z$ basis and one of the eigenstates each in $X$) are enough to secure the BB84 protocol, and Wang et al. \cite{PhysRevA.92.042319} showed that the RFI QKD protocol can be fully secured using only four states (two eigenstates of $Z$ basis and one of the eigenstates each in $X$ and $Y$ bases), and the resulting secret key rate is exactly the same as the original RFI QKD protocol. Hence, it still needs one more state compared with the BB84 protocol.

Recently, by adopting the method of convex optimization to estimate the Eve's information, Islam et al. \cite{PhysRevA.97.042347} reduced the number of the state in arbitrary $d$-dimensional QKD system. They show that the protocol can be secure even when using just one monitoring-basis state as long as the channel noise is low enough. In this paper, under the assumption of basis-independent state preparation, we find the number of states required in the RFI QKD protocol can be further reduced to three states (two eigenstates of $Z$ basis plus one of the eigenstates in $X$ basis) by using the semidefinite programming (SDP), and the secret key rate and transmission distance is still comparable to the original RFI QKD protocol. We noted that Bob also need randomly choose one of the $X$, $Y$ and $Z$ bases to measure the states sent from Alice, just like that in the original RFI QKD protocol. 

In the following of this paper, we first analyze the security for the generic RFI QKD protocol against arbitrary collective attacks, and then show that this protocol can be secure when using fewer states. An experimental demonstration based on time-bin encoding is proceeded to show the feasibility of our scheme, and its merit to simplify the experimental setup. In the experiment, we also consider the finite-key security against coherent attacks.

\section{Protocol and security}
We first briefly review the RFI QKD protocol\cite{Laing2010Reference}. It denotes the three Pauli matrices written $\left\{ {{\sigma _x},{\sigma _y},{\sigma _z}} \right\}$ by $\left\{ {X,Y,Z} \right\}$, and assume the one direction is well defined, i.e., ${Z_A} = {Z_B}$. The other two direction are allowed to change slowly in the quantum channel, that is , ${X_B} = \cos \beta {X_A} + \sin \beta {Y_A}$ and ${Y_B} = \cos \beta {Y_A} - \sin \beta {X_A}$. The meaning of $\beta $ depends on specific systems, such as the phase drift between Alice and Bob in time-bin encoding protocol. Besides, $\beta $ is unknown and may vary in time.

In each run, Alice (Bob) selects independently one of the three bases to prepare (measure) the quantum state. At the end of key distribution, They announce their bases. The raw keys are distilled from the events when they both use the well-defined $Z$ basis. The quantum bit error rate (QBER) is given by
\begin{eqnarray}
{e_{ZZ}} = \frac{{1 - \left\langle {{Z_A}{Z_B}} \right\rangle }}{2}
\label{e1}.
\end{eqnarray}
According the information collected on the bases complementary to $Z$, Alice and Bob can utilize an intermediate quantity $C$ to estimate Eve's knowledge. This quantity is defined as
\begin{eqnarray}
C = {\left\langle {{X_A}{X_B}} \right\rangle ^2} + {\left\langle {{X_A}{Y_B}} \right\rangle ^2} + {\left\langle {{Y_A}{X_B}} \right\rangle ^2} + {\left\langle {{Y_A}{Y_B}} \right\rangle ^2}
\label{e2}.
\end{eqnarray}
Here, note that $C$ is independent of relative angle $\beta$ when plugging the relations ${X_B}$ and ${Y_B}$ mentioned above into Eq.~(\ref{e2}). The maximal value under Pauli algebra is $C=2$, in this case, $e_{ZZ}=0$ as well: the two parameters $C$ and $e_{ZZ}$ are not independent, as we shall see in more detail later. When ${e_{ZZ}} \le 15.9\% $, Eve's information is given by
\begin{eqnarray}
{I_E} = \left( {1 - {e_{ZZ}}} \right)h\left( {\frac{{1 + \mu }}{2}} \right) + {e_{ZZ}}h\left( {\frac{{1 + \nu \left( \mu  \right)}}{2}} \right)
\label{e3},
\end{eqnarray}
where $h(x)$ is the binary Shannon entropy, and 
\begin{eqnarray}
\begin{array}{l}
\mu  = \min \left[ {\frac{{\sqrt {{C \mathord{\left/
 {\vphantom {C 2}} \right.
 \kern-\nulldelimiterspace} 2}} }}{{1 - {e_{ZZ}}}},1} \right],\\
\nu  = {{\sqrt {C/2 - {{\left( {1 - {e_{ZZ}}} \right)}^2}{\mu ^2}} } \mathord{\left/
 {\vphantom {{\sqrt {C/2 - {{\left( {1 - {e_{ZZ}}} \right)}^2}{\mu ^2}} } {{e_{ZZ}}}}} \right.
 \kern-\nulldelimiterspace} {{e_{ZZ}}}}.
\end{array}
\label{e4}
\end{eqnarray}

We present this protocol in an equivalent entanglement-based version where Alice and Bob share the state of the form ${\left| \phi  \right\rangle _{AB}} = \frac{1}{{\sqrt 2 }}\left( {{{\left| 0 \right\rangle }_A}{{\left| 0 \right\rangle }_B} + {{\left| 1 \right\rangle }_A}{{\left| 1 \right\rangle }_B}} \right)$. They independently implement a projective measurement on the entangled state to determine the state received by the counterpart. This fact, together with the assumption that it is possible to deal with finite-dimensional systems, guarantees that the security of the RFI QKD protocol can be analyzed against arbitrary collective attacks \cite{renner2005information}. Thus, each pair shared by Alice and Bob is supposed to be in two-qubit state ${\rho _{AB}}$, of which Eve holds a purification ${\rho _E} = {{\mathop{\rm Tr}\nolimits} _{AB}}\left( {{\rho _{ABE}}} \right)$, where ${{\rho _{ABE}}}$ is the density matrix shared among Alice, Bob and Eve after the transmission, and the state is ${\left| \Psi  \right\rangle _{ABE}} = \sum\nolimits_j {\sqrt {{\lambda _j}} } {\left| \phi  \right\rangle _{AB}}\left| {{E_j}} \right\rangle $, where $\left\langle {{E_i}} \right.\left| {{E_j}} \right\rangle  = {\delta _{ij}}$ is the orthogonal basis of system possessed by Eve.

In RFI QKD, it is obvious that the key ingredient is trying to obtain an optimal lower bound of $C$. We cast it into a minimization-SDP problem \cite{PhysRevA.97.042347,coles2016numerical}, where we use a priori known statistics of the compatible positive-operator value measure (POVM) of Alice and Bob, and measurement statistics extracted from the experiment. In this case, the problem can be converted into getting a well lower bound of ${C^L}$ under the following constraints \cite{PhysRevA.97.042347}
\begin{eqnarray}
{\mathop{\rm Tr}\nolimits} \left( {{{\hat E}_{ZZ}}{\rho _{AB}}} \right) = {e_{ZZ}},
\label{e5}
\end{eqnarray}
\begin{eqnarray}
{\mathop{\rm Tr}\nolimits} \left( {\hat P_{{\alpha _i}}^A \otimes \hat P_{{\chi _j}}^B{\rho _{AB}}} \right) = {P_{{\alpha _i},{\chi _j}}},
\label{e6}
\end{eqnarray}
\begin{eqnarray}
{\mathop{\rm Tr}\nolimits} \left( {{\rho _{AB}}} \right) = 1,
\label{e7}
\end{eqnarray}
\begin{eqnarray}
{\rho _{AB}} \ge 0,
\label{e8}
\end{eqnarray}
where $\forall \left\{ {\alpha ,\chi } \right\} \in \left\{ {X,Y,Z} \right\}$ and $\forall \left\{ {i,j} \right\} \in \left\{ {0,1} \right\}$. For four-state scheme, it is noted that $i = 0$ if $\alpha  = X$ or $Y$. When three-state scheme is adopted, $\alpha  \ne Y$ and $i = 0$ if $\alpha  = X$. ${\hat E_{ZZ}}$ and ${\hat P_i} = \left| i \right\rangle \left\langle i \right|$ are, respectively, the error operator in the $Z$ basis and the projective measurement on the entangled state $\rho _{AB}$. They are well defined according the protocol. The statistics of probability $P_{{\alpha _i},{\chi _j}}$ can be extracted from the experiment, which are the probabilities that Bob receiving a state $\left| {{\chi _j}} \right\rangle$ after Alice obtained a state $\left| {{\alpha _i}} \right\rangle $ by measuring her photon. Here, the $\rho _{AB}$ is allowed to be arbitrary, which means Eve can perform any operations on the states transmitted between Alice and Bob, and hence the bound is valid for any collective attacks respecting the given measurement statistics \cite{PhysRevA.97.042347}. This optimization problem can be efficiently solved by using Matlab package CVX designed for disciplined convex programming \cite{cvx,gb08}. The detail of calculations are shown in Appendix A.

\begin{figure}[htbp]
\centering
\includegraphics[width=0.48\textwidth]{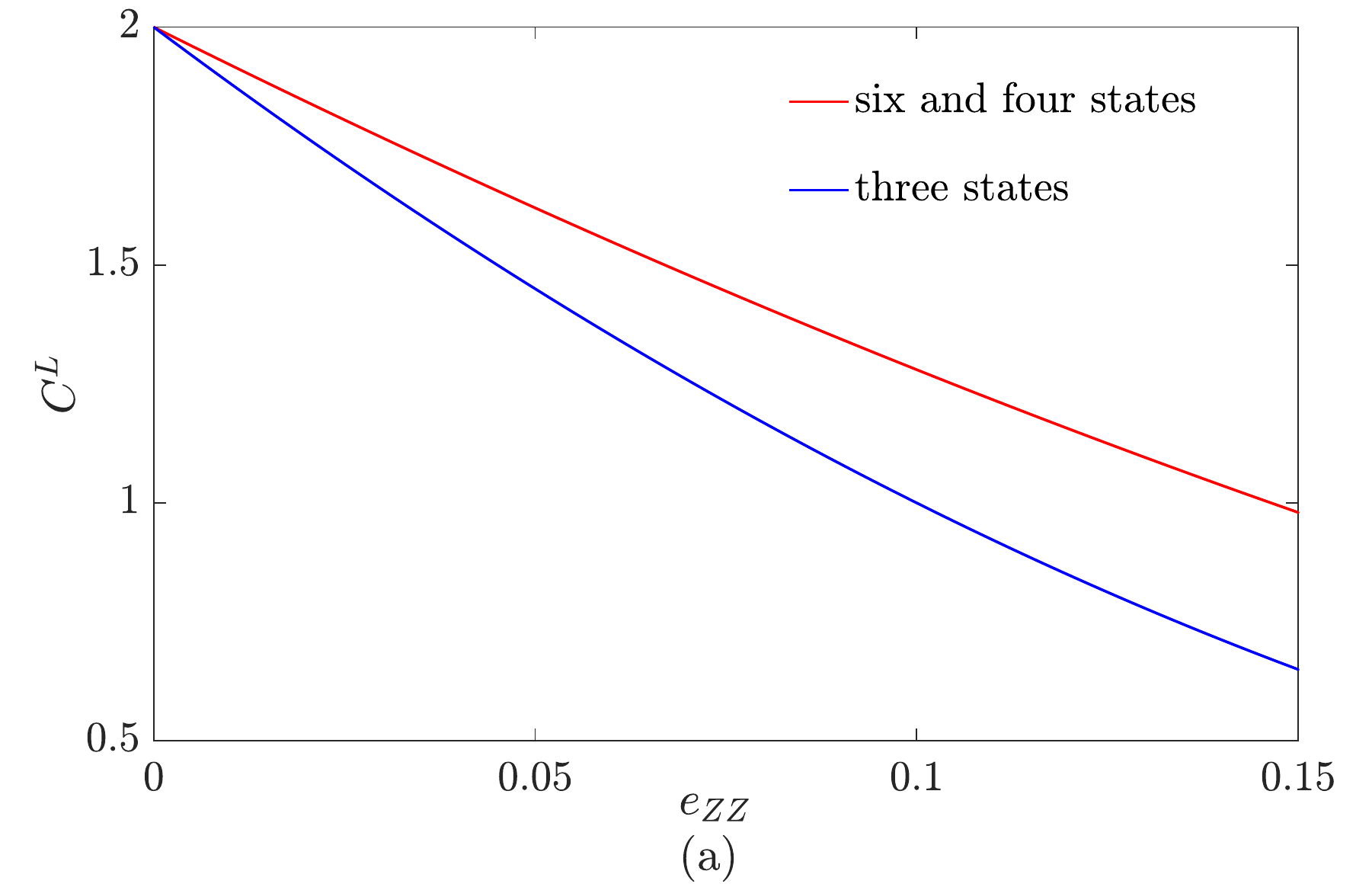} \\
\includegraphics[width=0.48\textwidth]{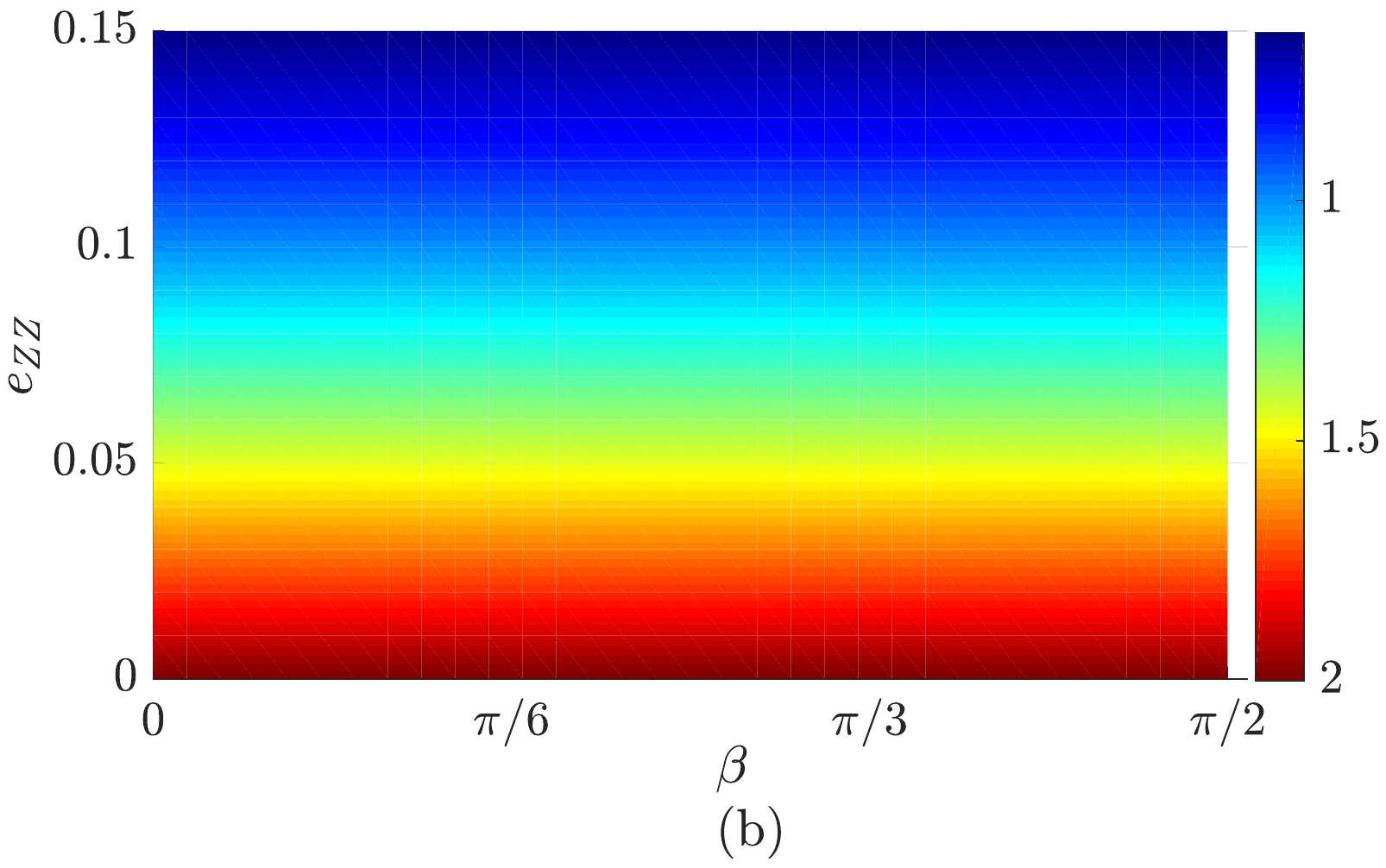}
\caption{\label{fig1}(Color online) (a) The numerically obtained lower bounds of the $C_L$ plotted as a function of the quantum bit error rates $e_{ZZ}$ for the case where Alice sends six, four, and three states respectively. (b) The lower bounds of $C_L$ as a function of the quantum bit error rates $e_{ZZ}$ and misalignments of the reference frame $\beta$ when only three states are sent.}
\end{figure}

To demonstrate our scheme, we treat the bit-flip rates in the $X$, $Y$ and $Z$ bases equivalently for simplicity, i.e., ${e_z} = {e_x} = {e_y} = {e_{ZZ}}$, which are caused by channel noise in the transmission. In the original RFI QKD protocol, Alice needs to randomly prepare six states ${\left| Z_0 \right\rangle },{\left| Z_1 \right\rangle },{\left| X_0 \right\rangle },{\left| X_1 \right\rangle },{\left| Y_0 \right\rangle }$, and ${\left| Y_1 \right\rangle }$. By using the method  proposed in Ref.~\cite{PhysRevA.92.042319}, the number of states can be reduced to four, where states ${\left| X_1 \right\rangle }$ and ${\left| Y_1 \right\rangle }$ are redundant to estimate $C_L$. It can be seen from Fig.~\ref{fig1}(a) that the method of optimization applied here correspond to this result, as $C_L$ is the same when Alice sends only four states in comparison to the case when she send all six states (red solid line). Furthermore, our SDP approach is available to estimate the lower bound of $C$ when only three states are sent by Alice, as shown by the solid blue line in Fig.~\ref{fig1}(a). Note that it is still required for Bob to randomly pick one of three bases $X$, $Y$ and $Z$ to measure Alice's three states ${\left| Z_0 \right\rangle },{\left| Z_1 \right\rangle }$ and ${\left| X_0 \right\rangle }$. In the four-state scenario, complete knowledge of the remaining unused states can be reconstructed from the statistics of four used states and from the statistics of the events where Alice and Bob choose different basis. However, when Alice sends only three states, complete knowledge of the non-transmitted states cannot be reconstructed using the experimentally determined statistics. Thus, the lower bound of $C$ decreased faster as $e_{ZZ}$ increased. Nonetheless, $C_L$ at the three-state scenario is still independent to the change of $\beta$, as shown in Fig.~\ref{fig1}(b), This validates that our scheme is reference frame independent, even if only $X$ and $Z$ bases are employed to prepare encoded states.

\section{Estimation of the secret key rates}
After obtaining the value of $C$, in this section, we estimate the key generation rate of RFI QKD using different number of states based on two kinds of source: single-photon source and phase-randomized weak coherent source (WCS) with vacuum+weak decoy state.
\subsection{Single-photon source}
In the original six states RFI QKD protocol, the calculation of $C$ need applies the equivalent formula of Eq.~(\ref{e2}), it can be written as \cite{yin2014reference}
\begin{eqnarray}
\begin{aligned}
C' = &{\left( {1 - 2{e_{XX}}} \right)^2} + {\left( {1 - 2{e_{XY}}} \right)^2}\\
 & + {\left( {1 - 2{e_{YX}}} \right)^2} + {\left( {1 - 2{e_{YY}}} \right)^2},
\end{aligned}
\label{e9}
\end{eqnarray}
where ${e_{\alpha \chi }}$ is the phase error rate defined as a fictitious bit error rate in the $\alpha$ basis and the $\chi$ basis. It is a virtual procedure that Alice first prepares an entanglement state in the $Z$ basis and then Alice (Bob) measure it in the $\alpha$ ($\chi$) basis. If there are no basis-dependent source flaws, the ${e_{\alpha \chi }}$ equals the QBER when Alice prepares her state in the $\alpha$ basis, and Bob measures it in the $\chi$ basis, which can be directly measured in experiments. For our three states scheme, only ${e_{ZZ}}$, ${e_{XX}}$, and ${e_{XY}}$ can be extracted from the experiment, these parameters as prior known statistics are sufficient to  estimate the lower bound of $C$, since the fictious bit error rates $e_{YX}$ and $e_{YY}$ can be well bounded using the SDP method, as shown in Appendix A. After obtaining $C$ and $e_{ZZ}$, we can estimate Eve's information ${I_E}$ in Eq.~(\ref{e3}) for the RFI QKD protocol.

\begin{figure}[htbp]
\centering
\includegraphics[width=0.48\textwidth]{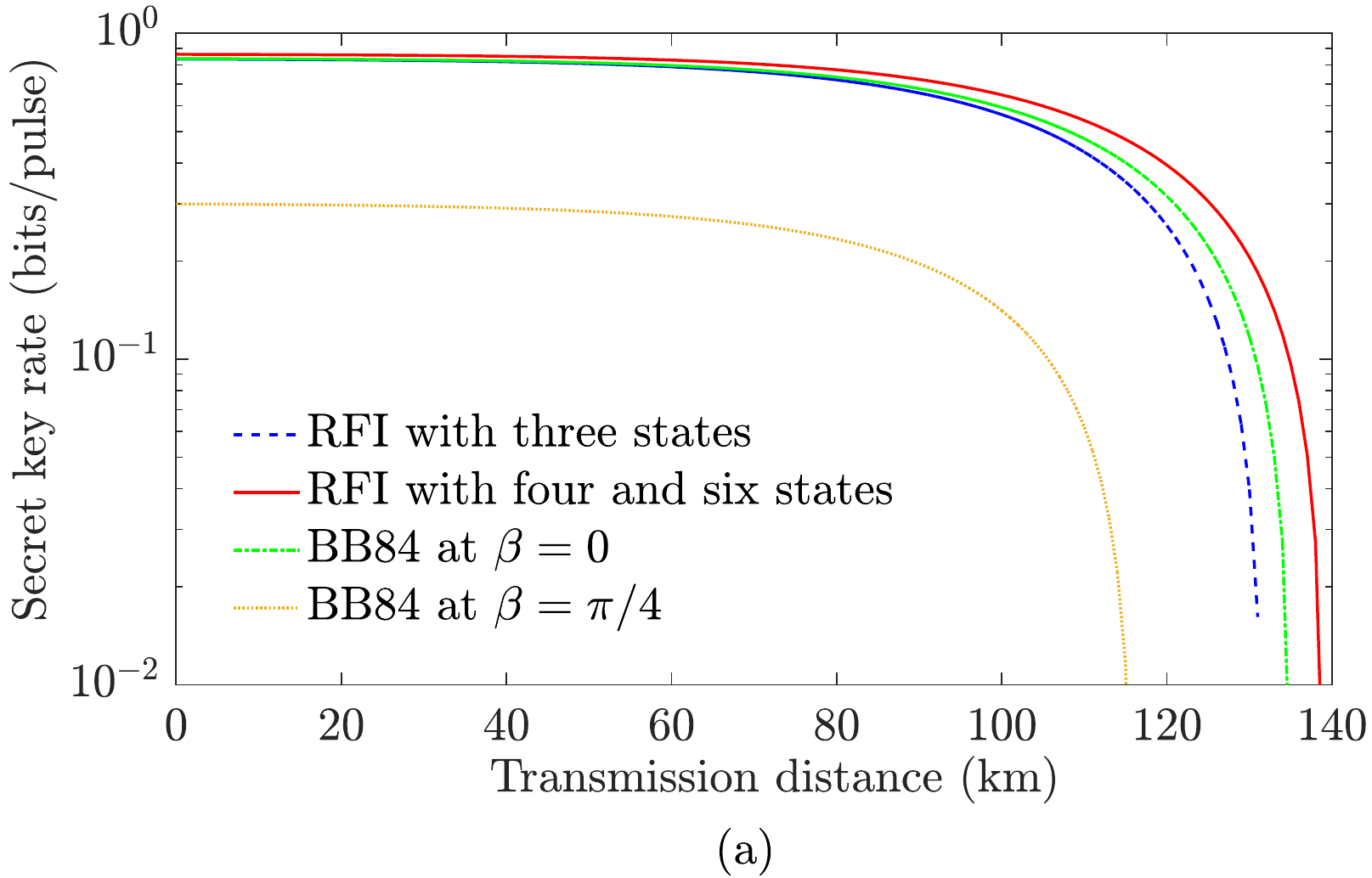} \\
\includegraphics[width=0.48\textwidth]{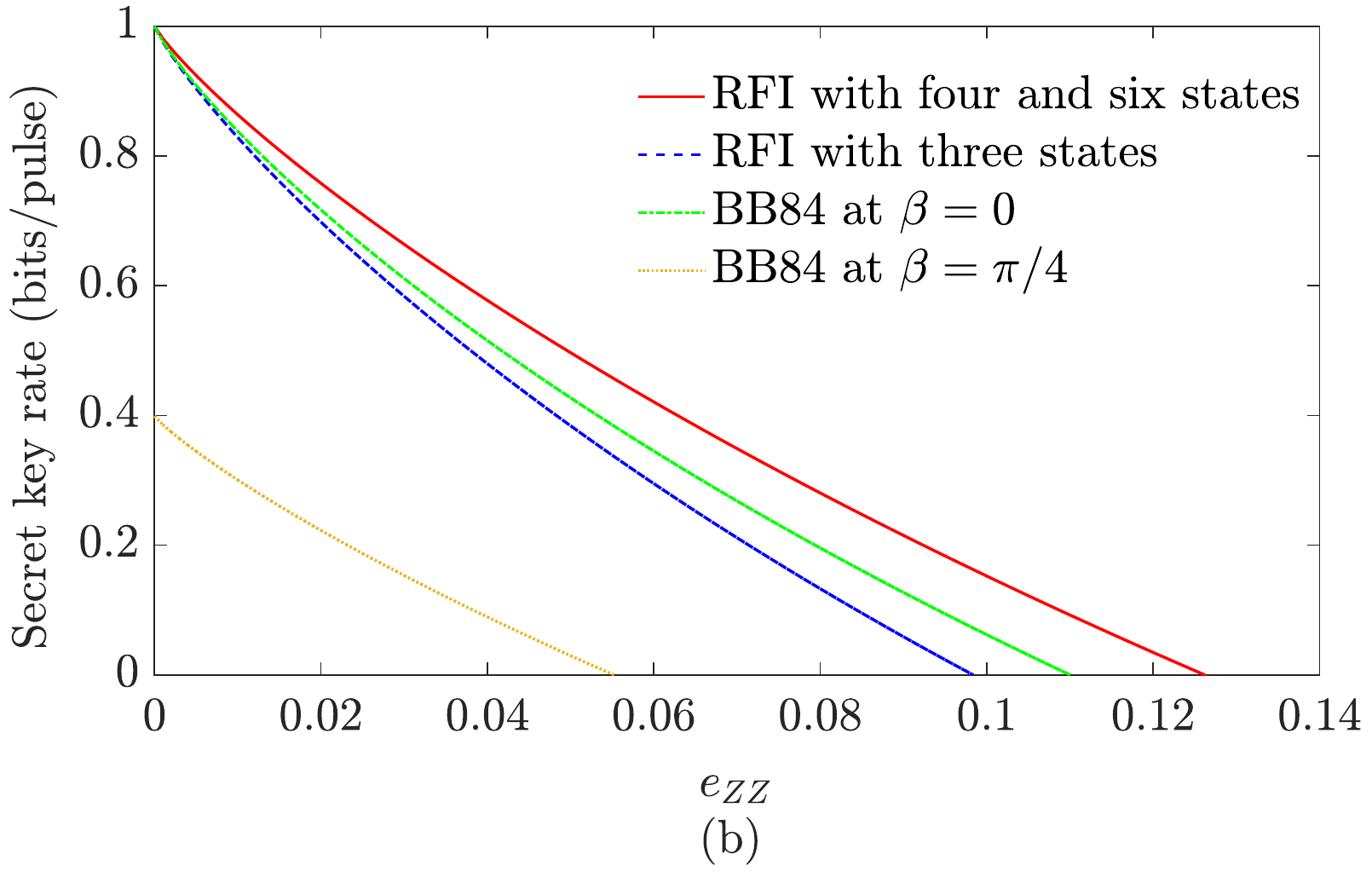} 
\caption{\label{fig2}(Color online) In the case of single-photon source employed, the secret key rates as a function of the transmission distances (a) and the quantum bit error rates (b) for the RFI QKD protocol and the BB84 protocol at the different misalignment of reference frame.}
\end{figure}

In an asymptotic case, the secret key rate defined as the number of bits per pulse is given by
\begin{eqnarray}
R = 1 - h\left( {{e_{ZZ}}} \right) - {I_E}.
\label{e10}
\end{eqnarray}
By consider the channel model proposed in Ref.~\cite{PhysRevA.90.052314}, we simulate the secret key rates along with the change of the transmission distances and $e_{ZZ}$ as shown in Fig.~\ref{fig2}(a) and Fig.~\ref{fig2}(b) respectively. For comparison, we also simulate the key generation rate of three states BB84 protocol using the SDP approach \cite{PhysRevA.97.042347}. When the state-preparation process is assumed ideal, Eve's information is estimated as ${I_E} = h\left( {{e_{XX}}} \right)$ for the BB84 protocol. The parameters in our method are set according our experiment system, i.e., the dark count rate of single photon detector (SPD) is ${e_d} = 8 \times {10^{ - 6}}$, the detection efficiency is ${\eta _d} = 13\% $, the optical intrinsic error rate is $e_o=0.01$ and the loss coefficient of the channel is 0.21 dB/km.

For the specific case where Alice sends all six states (Fig.~\ref{fig2}(b), solid red line), it is evident that the maximum error tolerance is $ \sim 12.6\% $, which is in agreement with existing findings \cite{Laing2010Reference}. The error tolerance is defined as the error rate $e_{ZZ}$ beyond which $R=0$. Furthermore, we find that secret key rates are identical when Alice sends only four states (Fig.~\ref{fig2}(a), solid red line) in comparison to the case when she sends all six states, illustrating that two of the states are redundant. For the case where Alice sends only three states, $e_{XX}$ and $e_{XY}$ can be substitute into our SDP method (shown in Appendix A) to estimate a low bound of $C_L$. It is evident that the three states RFI QKD protocol still generates a positive secret key rate, as illustrated by blue dashed line in Fig.~\ref{fig2}, but with a lower error tolerance ($ \sim 9.8\% $). Despite the lower error tolerance, we observe that the maximal transmission distance of the three states RFI QKD protocol is close to that of the RFI QKD with four and six states and that of the BB84 protocol when no misalignment of reference frame occurs. Moreover, at the distances of less than 80 km, it is evident that their curves are almost overlapped. In the case of reference frame misalignment, the secret key rates of the RFI QKD protocol remain the same at different $\beta$ whenever six states or three states are sent by Alice. However, the transmission distance and the secret key rate of the BB84 protocol are decreased dramatically at $\beta=\pi/4$, as shown dotted yellow line in Fig.~\ref{fig2}(a). These results verify that our scheme based on the single-photon source can release the requirement of the calibration of reference frame, even when only three states are prepared by Alice. We noted that Bob still need pick one of the $X$, $Y$, and $Z$ bases to measure the states sent from Alice, just like that in the original RFI QKD protocol.

\subsection{Phase-randomized WCS with decoy-state method}
The secret key rate calculated above is based on an ideal single-photon source. However, in most practical QKD systems, a phase-randomized WCS combined with decoy-state method is generally employed to overcome the photon-number-splitting (PNS) attack against the multiphoton pulses \cite{DECOY05,Wang2005Beating,PhysRevA.89.022307}. In particular, we assume that Alice can set the intensity of each laser pulse to one of the three predetermined intensity levels, $\mathcal{K} \in \left\{ {\mu ,\nu ,\omega } \right\}$, each transmitted with a probability $p_k$. Three intensity levels satisfy the conditions: $\mu  > \nu  + \omega $ and $0 \le \omega  \le \nu $. Furthermore, in the applications, the number of total pulses $N$ sent by Alice is always finite; thus, we must consider the effect of statistical fluctuation caused by a finite size fo pulses to ensure the security of the RFI QKD protocol.

In such a scenario, we consider the number of pulses sent by Alice to be $N = {10^{10}}$. The probability of Alice (Bob) preparing (measuring) a state with $\alpha$ basis is $\Pr_\alpha ^{A\left( B \right)}$. Here, we use all intensity levels for the key generation. According to Ref.~\cite{sheridan2010finite,Zhang:17}, the secret key length against coherent attack is
\begin{eqnarray}
\begin{aligned}
\ell  =& \left\lfloor {{s_{zz,0}} + {s_{zz,1}}\left( {1 - {I_E}} \right) - {n_{zz}}fh\left( {{E_{ZZ}}} \right)} \right. \\
 &\left. { - {{\log }_2}\frac{2}{{{\varepsilon _{EC}}}} - 2{{\log }_2}\frac{1}{{{\varepsilon _{PA}}}}} \right.\\
&\left. { - 7{n_{zz}}\sqrt {\frac{{{{\log }_2}\left( {{2 \mathord{\left/
 {\vphantom {2 {\bar \varepsilon }}} \right.
 \kern-\nulldelimiterspace} {\bar \varepsilon }}} \right)}}{{{n_{zz}}}}}  - 30{{\log }_2}\left( {N + 1} \right)}\right\rfloor,
\label{e11}
\end{aligned}
\end{eqnarray}
where $s_{zz,0}$ and $s_{zz,1}$ are the number of vacuum events and the number of single-photon events associated with the single-photon events in $Z_A$ respectively. $E_{ZZ}$ is the average of the observed error rate in basis $Z$, $f$ denotes the inefficiency of error correction, $n_{zz}$ is the number of detected pulses when Alice prepares her state in the $Z$ basis and Bob measures it in the $Z$ basis. ${{\varepsilon _{EC}}}$ (${\varepsilon _{PA}}$) denotes the probability that error correction (privacy amplification) fails, and ${\bar \varepsilon }$ measures the accuracy of estimating the smooth min-entropy \cite{sheridan2010finite}. In this paper, we set ${\varepsilon _{EC}} = {\varepsilon _{PA}} = \bar \varepsilon =\varepsilon = {10^{ - 10}}$.

\begin{figure}[htbp]
\centering
\includegraphics[width=0.48\textwidth]{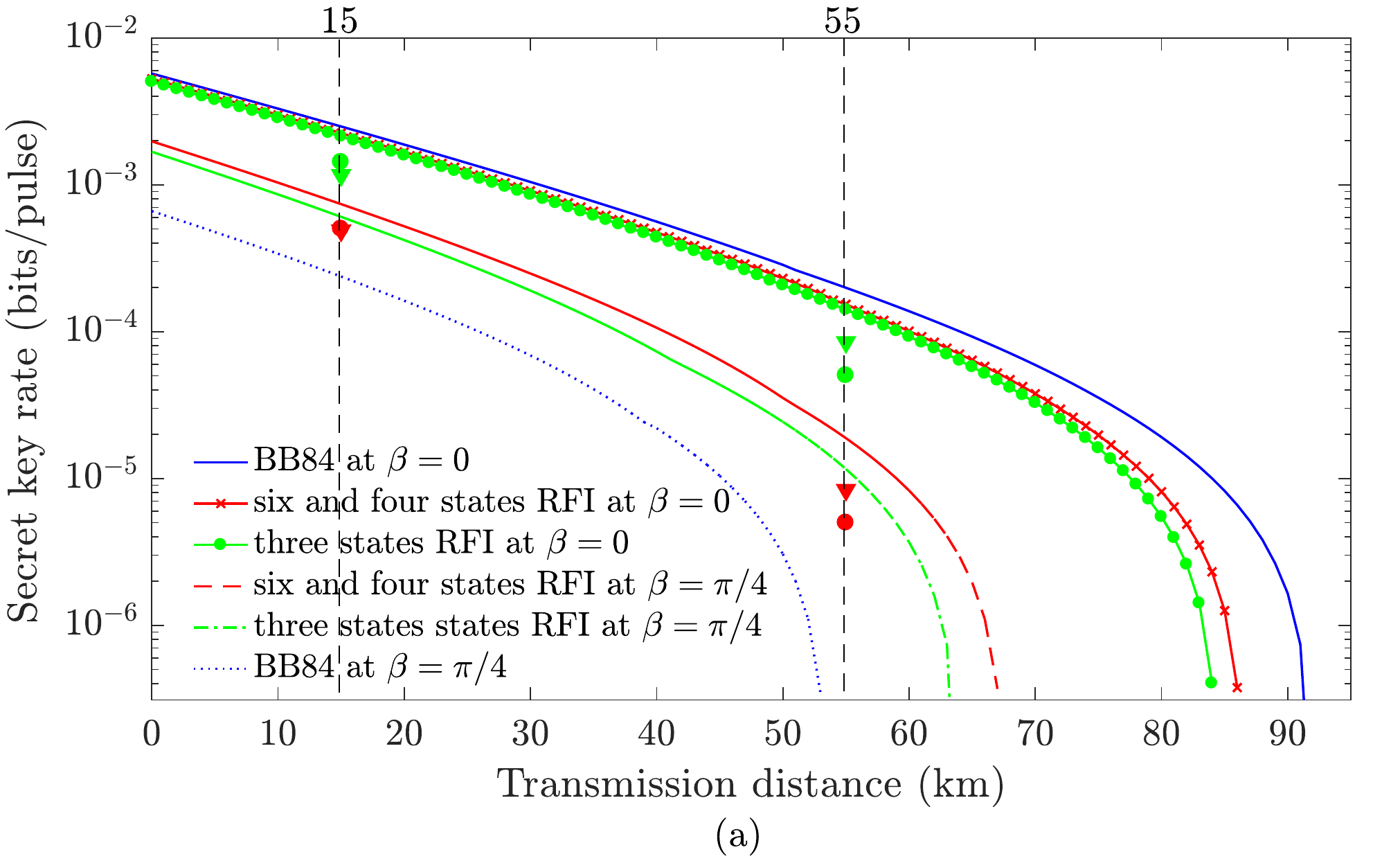} \\
\includegraphics[width=0.48\textwidth]{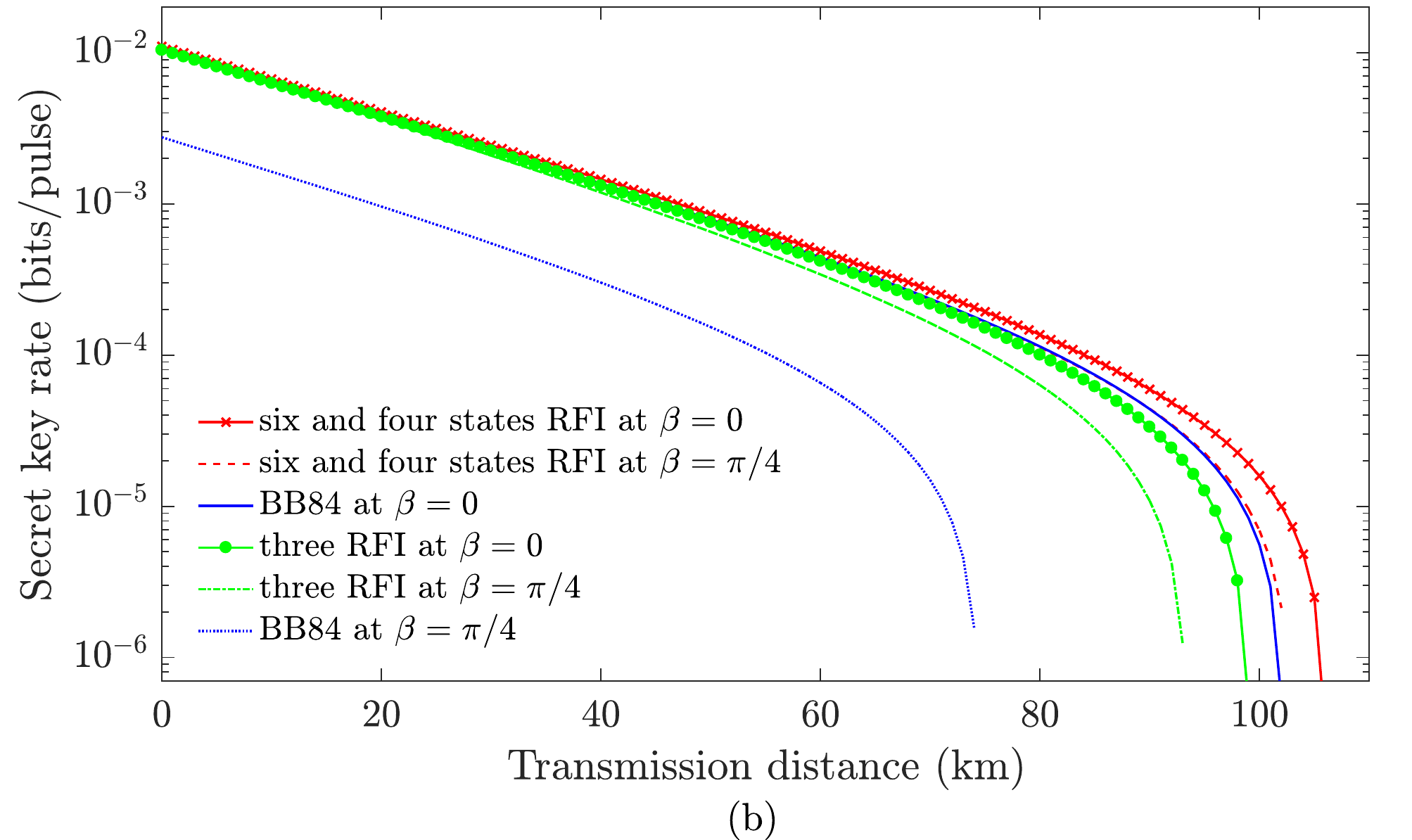} 
\caption{\label{fig3}(Color online) Secret key rate vs. transmission distances when the weak coherent source is used. (a) Considering the statistical fluctuations dut to the finite-size pulses, numerically optimized secret key rates against coherent attacks are obtained for a fixed postprocessing block size $N = {10^{10}}$. The dot symbols are the experimental results when Alice sends three states to Bob, and the triangle markers are the results for the four-states protocol. The green color and the red color respectively represent $\beta {\rm{ = 0}}$ and $\beta {\rm{ = \pi/4}}$. (b) The secret key rates in the asymptotic case, i.e., in the limit of infinitely large keys.}
\end{figure}

Applying the method proposed in Ref.~\cite{PhysRevA.89.022307}, the number of vacuum events in $Z_A$ satisfies
\begin{eqnarray}
{s_{zz,0}} \ge \max \left[ {{\tau _0}\frac{{\left( {\nu n_{zz,\omega }^ -  - \omega n_{zz,\nu }^ + } \right)}}{{\nu  - \omega }},0} \right],
\label{e12}
\end{eqnarray}
where ${\tau _n}: = \sum\nolimits_{k \in \mathcal{K}} {{{{e^{ - k}}{k^n}{p_k}} \mathord{\left/ {\vphantom {{{e^{ - k}}{k^n}{p_k}} {n!}}} \right. \kern-\nulldelimiterspace} {n!}}} $ is the probability that Alice sends a $n$-photon state, and
\begin{eqnarray}
n_{zz,k}^ \pm : = \max \left[ {\frac{{{e^k}}}{{{p_k}}}\left( {{n_{zz,k}} \pm \sqrt {\frac{{{n_{zz}}}}{2}\ln \frac{1}{\varepsilon }} } \right),0} \right]
\label{e13}
\end{eqnarray}
The number of single-photon events in $Z_A$ is
\begin{eqnarray}
\begin{aligned}
{s_{zz,1}} \ge &\max \left\{{\frac{{{\tau _1}\mu }}{{\mu \left( {\nu  - \omega } \right) - {\nu ^2} + {\omega ^2}}}} [ {n_{zz,\nu }^ -  - n_{zz,\omega }^ + }\right. \\
 &\left. {{ - \frac{{{\nu ^2} - {\omega ^2}}}{{{\mu ^2}}}\left( {n_{zz,\mu }^ +  - {{{s_{zz,0}}} \mathord{\left/
 {\vphantom {{{s_{zz,0}}} {{\tau _0}}}} \right.
 \kern-\nulldelimiterspace} {{\tau _0}}}} \right)}], 0} \right\}.
\label{e14}
\end{aligned}
\end{eqnarray}
The QBER ${e_{ZZ}}$ associated with the single-photon events in $Z_A$ is given by
\begin{eqnarray}
{e_{ZZ}} \le \min \left[ {{\tau _1}\frac{{m_{zz,\nu }^ +  - m_{zz,\omega }^ - }}{{\left( {\nu  - \omega } \right){s_{zz,1}}}},\frac{1}{2}} \right],
\label{e15}
\end{eqnarray}
where
\begin{eqnarray}
m_{zz,k}^ \pm : = \max \left[ {\frac{{{e^k}}}{{{p_k}}}\left( {{m_{zz,k}} \pm \sqrt {\frac{{{m_{zz}}}}{2}\ln \frac{1}{\varepsilon }} } \right)},0 \right].
\label{e16}
\end{eqnarray}
We also calculate the number of vacuum events, ${s_{\kappa \zeta ,0}}$, and the number of single-photon events,  ${s_{\kappa \zeta ,1}}$, for $\kappa  = { \cup _{k \in \mathcal{K}}}{\kappa _k}$, where ${\forall \left\{ {\kappa ,\zeta } \right\} \in \left\{ {X,Y} \right\}}$, i.e., by using Eqs.~(\ref{e12}) and (\ref{e14}) with statistics from the basis $\kappa$. In addition, the formula for the phase error rate of single-photon events in ${\kappa_A} {\zeta_B}$ is \cite{PhysRevA.97.040301}
\begin{eqnarray}
{e_{\kappa  \zeta  }} \le \min \left\{ {\left[ {{{\tilde e}_{\kappa \zeta  }} + \gamma \left( {\varepsilon ,{{\tilde e}_{\kappa  \zeta  }},{s_{\kappa \zeta  ,1}},{s_{zz,1}}} \right)} \right],\frac{1}{2}} \right\},
\label{e17}
\end{eqnarray}
where ${{{\tilde e}_{\kappa \zeta  }}}$ can be calculated using Eq.~(\ref{e15}), and
\begin{equation}
\gamma \left( {a,b,c,d} \right): = \sqrt {\frac{{\left( {c + d} \right)\left( {1 - b} \right)b}}{{cd}}\ln \left[ {\frac{{c + d}}{{2\pi cd\left( {1 - b} \right)b{a^2}}}} \right]}.
\label{e18}
\end{equation}

For the evaluation, we numerically optimize the secret key rate $R: = {\ell  \mathord{\left/ {\vphantom {\ell  N}} \right. \kern-\nulldelimiterspace} N}$ over the free parameters $\left\{ {\Pr _Z^A,{p_u},{p_v},\mu,\nu} \right\}$ as shown in Fig.~\ref{fig3}(a). Due to the symmetry of the $X$, $Y$ basis in Eq.~(\ref{e2}), we treat the parameters of the these two bases equivalently for simplicity. Accordingly, $\Pr _X^A = \Pr _X^B = \Pr _Y^A = \Pr _Y^B$, expect for $\Pr _Y^A=0$ as Alice sent three states to Bob. According to the experiment system, we set $\omega  = 0$ to be a vacuum state. For purposes of comparison, the secret key rates in the asymptotic case are simulated as shown in Fig.~\ref{fig3}(b), and the secret key rates for the BB84 protocol are also depicted in Fig.~\ref{fig3} by using the blue solid line and the blue dotted line. In the finite-key case for the BB84 protocol, we adopt the formula proposed in Ref.~\cite{PhysRevA.89.022307} to estimate its secret key rates. For the RFI QKD protocol, the achieved key rates will be the lowest in the finite-key case at $\beta=\pi/4$, which can be explained by poor estimation of $C$ with the increase of $\beta$. It is evident that the secret key rates of the RFI QKD protocol are still an order of magnitude higher compared with that of the BB84 protocol at $\beta=\pi/4$. For the case where Alice only sends three states, the phase error rates $e_{XX}$ and $e_{XY}$ can be estimated according to Eq.~(\ref{e17}), and they are then taken into SDP approach to get the value of $C_L$. We show that the secret key rate and the transmission distance are comparable with that of the original six states RFI QKD protocol, which verify the feasibility of our scheme in the real-world applications.

\section{Experimental setup and results}
\begin{figure}[htbp]
\centering
\includegraphics[width=0.48\textwidth]{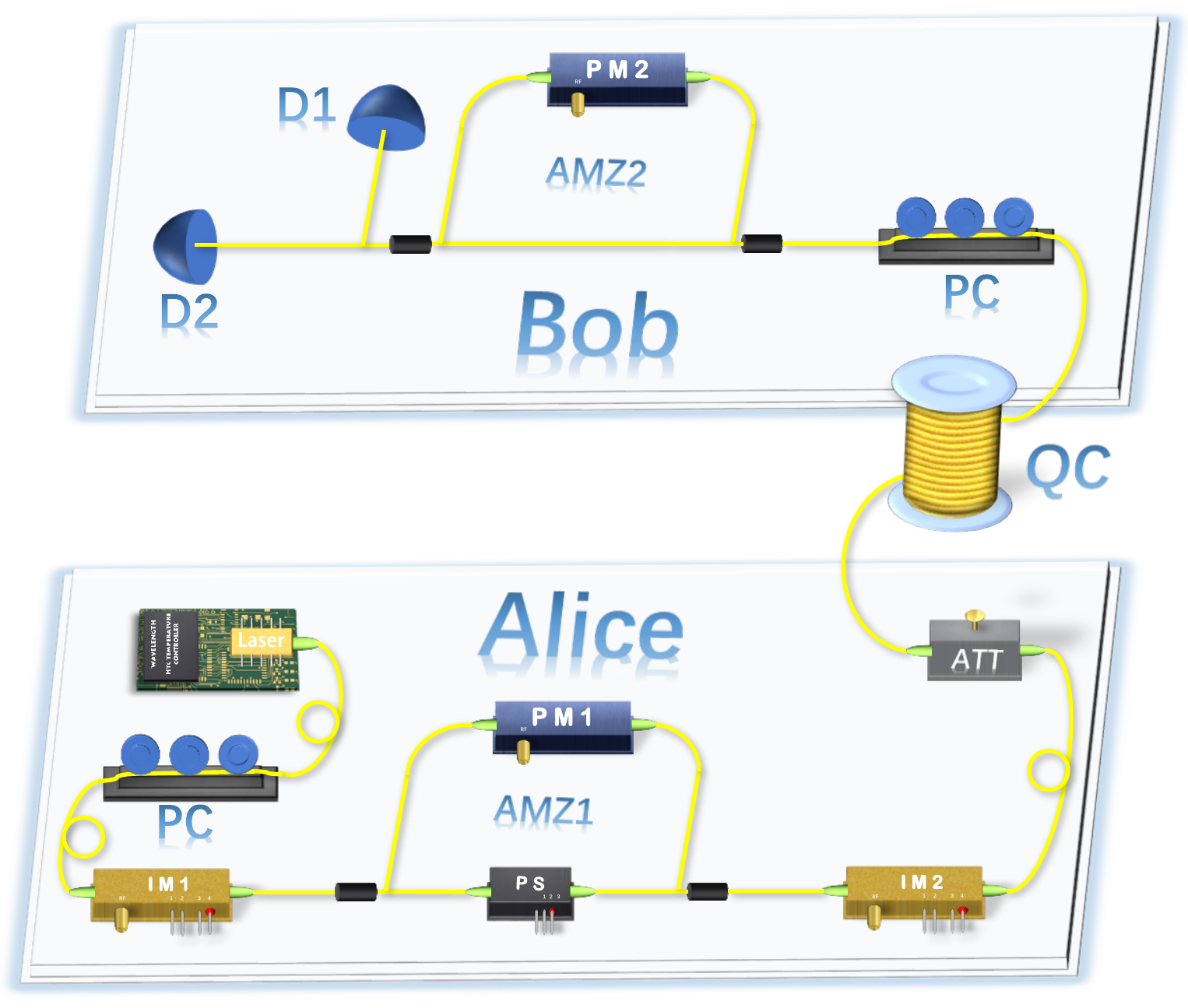}
\caption{\label{fig4}(Color online) Schematic of the experiment. PC, polarization controller; IM, intensity modulator; PM, phase modulator; PS, phase shifter; ATT, attenuator; D1,D2, single-photon detector (SPD); QC, an SMF-28 fiber spool, which has a channel attenuation of $\alpha {\rm{ = 0}}{\rm{.21dB/km}}$.}
\end{figure}

\begin{table*}
\caption{\label{t1}Implementation parameters and experimental results.}
\begin{ruledtabular}
\begin{tabular}{ccccccccccccccc}
  \multirow{3}{*}{Protocol} & \multirow{3}{*}{$\beta$} &\multicolumn{5}{c}{Parameters}&\multicolumn{6}{c}{Estimation} & \multicolumn{2}{c}{Performance}\\ \\
& & $\mu$ & $\nu$ & $p_{\mu}$ & $p_{\nu}$ & $\Pr _Z^A$ & $s_{zz,0}$ & $s_{zz,1}$ & $e_{ZZ}$ & $e_{XX}$ & $e_{XY}$ & $C_L$ & $E_{ZZ}$ & $R$ \\
\hline \\
\multicolumn{2}{c}{15 km  7.95 dB} & & & & & & & & & & & & &\\
\cline{1-2}\\
Four & $0$ &0.59 &0.26 &0.59 &0.32 &0.89 & 0 & $1.78 \times 10^7$& 0.83\% & 4.60\%& 50\% & 1.68 & 0.44\% & $1.16 \times 10^{-3}$\\
Three & $0$ &0.58 &0.25 &0.60 &0.31 &0.90 & 0 & $1.97 \times 10^7$& 0.72\% & 2.62\%& 50\% & 1.77 & 0.42\% & $1.42 \times 10^{-3}$\\
Four & $\pi/4$ & 0.47 & 0.14 & 0.43 & 0.39 & 0.79 & $2.18 \times 10^4$ & $1.07 \times 10^7$ & 0.97\% & 20.74\% & 21.15\% & 1.33 & 0.61\% & $4.88 \times 10^{-4}$\\
Three & $\pi/4$ & 0.44 & 0.12 & 0.44 & 0.39 & 0.78 & $1.83 \times 10^4$ & $1.00 \times 10^7$ & 0.70\% & 20.39\% & 19.44\% & 1.42 & 0.63\% & $5.00 \times 10^{-4}$\\
\hline \\
\multicolumn{2}{c}{55 km  16.35 dB} & & & & & & & & & & & & &\\
\cline{1-2}\\
Four & $0$ &0.59 &0.28 &0.36 &0.48 &0.82 & $4.88 \times 10^4$ & $2.21 \times 10^6$& 2.08\% & 6.21\%& 50\% & 1.50 & 1.89\% & $8.48 \times 10^{-5}$\\
Three & $0$ &0.55 &0.28 &0.38 &0.42 &0.81 & $7.47 \times 10^4$ & $2.00 \times 10^6$& 1.85\% & 8.20\%& 50\% & 1.34 & 2.30\% & $5.02 \times 10^{-5}$\\
Four & $\pi/4$ & 0.43 & 0.15 & 0.22 & 0.47 & 0.62 & $5.50 \times 10^4$ & $7.25 \times 10^5$ & 1.32\% & 23.36\% & 23.41\% & 1.13 & 3.90\% & $8.36 \times 10^{-6}$\\
Three & $\pi/4$ & 0.39 & 0.12 & 0.23 & 0.45 & 0.59 & $5.20 \times 10^4$ & $5.40 \times 10^5$ & 1.71\% & 21.76\% & 21.56\% & 1.23 & 4.25\% & $4.98 \times 10^{-6}$\\
\end{tabular}
\end{ruledtabular}
\end{table*}

To demonstrate the application of our scheme to a real QKD system, we proceed a proof-of-principle experiment using time-bin encoding, as shown in Fig.~\ref{fig4}. The light pulses generated by Alice's coherent light source (1550 nm) are randomly modulated into two intensities of decoy states using an intensity modulator (IM1). The vacuum state is generated by stopping the trigger on the laser. Then, the quantum states of photons are modulated by an asymmetric Mach-Zehnder interferometer (AMZ1) together with IM2 according to the coding information. For the $Z$ basis, the key bit is encoded in time bin, 0 or 1, by IM2. For the $X$ and $Y$ bases, the key bit is encoded into the relative phase, 0 or $\pi$ for $X$ basis, and $\pi/2$ or $3\pi/2$ for $Y$ basis, by phase modulator (PM1). A phase shifter (PS) in AMZ1 is used to simulate the change of the reference frame. After pulses passed through the AMZ1, the time interval of two adjacent pulses is 7 ns. The repetition rate of the system is set to 1 MHz using a digital waveform generator based on a field-programmable gate array (FPGA, not shown here for clarity). Light pulses are then attenuated to the single-photon level by a attenuator and transmitted through a quantum channel (QC) to Bob. 

To demodulate the information, Bob needs to make measurement of the arriving photons on a randomly and independently selected basis. There are three possible time-bins of the photons arriving at Bob's single photon detectors (SPD). When Bob chooses $Z$ basis to measure the received photons, the detectors D1 and D2 are respectively aligned at the first and the third time-bin. The PM2 is used to perform the $X$ or the $Y$ basis measurement, i.e., the phase 0 for $X$ basis measurement,  and $\pi/2$ for $Y$ basis measurement. In this case, D1 and D2 are aligned at the second time-bin.

In the experiment, a higher modulation phase means that PM requires a higher driven voltage. For the original six states RFI QKD protocol, three independent signals need to be combined at a $3 \times 1$ coupler and the output signal is then used to drive the PM. The discrepancy of the arrival times to the PM for these three signals will lead to state-dependent error rates in the $X$ and $Y$ bases. Even though only $\pi/2$ needs to modulate in the case where Alice sends four states to Bob, the inaccurate value of driven voltage can result in a imperfect phase value, which will weaken the security of the protocol \cite{PhysRevA.92.042319}. In our three states scheme, the above problems can be avoided, since the PM1 is redundant to prepare a state $\left| {{X_{\rm{0}}}} \right\rangle$, which undoubtedly can reduce the complexity of a real system. 

The interference visibility of our system is 97.2\%. Here, the experimental validation with the case where Alice sends four and three states to Bob is carried out at the transmission distances of 15 km and 55 km, and the different misalignments of the reference frame $\beta$ are considered. By plugging the experimental counts into the decoy-state estimations and using Eq.~(\ref{e11}), we obtain the experimental results listed in Table \ref{t1} and Fig.~\ref{fig3}(a). The estimations of $C_L$ are obtained by substituting the estimated error rate, i.e., $e_{ZZ}$, $e_{XX}$, $e_{XY}$, $e_{YY}$,and $e_{YX}$, into the SDP model. The deviations between the simulation results and the experimental results are primarily due to the excess loss of 3 dB when Bob measured the received states with the $Z$ basis. As expected, it is seen that the secret key rates are almost identical for the four-states and three-states scheme at the transmission distance of 15 km. The green and the red triangles in Fig.~\ref{fig3}(a) respectively denote the experimental results at $\beta=0$ and $\beta=\pi/4$ using the four-states scheme; the green and the red dots respectively represent the results of three-states scheme at $\beta=0$ and $\beta=\pi/4$. At the distance of 55 km, the secret key rates for four-states scheme (the triangle symbols) are slightly higher than that of three-states scheme (the dot symbols).

\section{Conclusion}
In summary, we propose an efficient scheme to realize the RFI QKD by using only three states, which is identical to the BB84 protocol. Furthermore, the secret key rates and the transmission distance are comparable to the original RFI QKD protocol. Experiments considering the finite-key analysis are demonstrated at the transmission distance of 15 km and 55 km. Our scheme is also suitable to the free-space RFI QKD systems, and can be upgraded to the RFI measurement-device-independent (MDI) QKD protocol \cite{Liu18} and high-dimension RFI QKD protocol \cite{liang2014proof} with simple modifications to the setup, thereby reducing the complexity of these systems.
\begin{acknowledgments}
This work was supported by the National Natural Science Foundation of China (NSFC) (11674397), and the Fund of State Key Laboratory of Information Photonics and Optical Communications (Beijing University of Posts and Telecommunications) (No. IPOC2017ZT04), P. R. China.
\end{acknowledgments}

\appendix

\section{Explicit calculation for $C_L$}
The estimation of lower bound of $C$ can be turned into an optimization framework \cite{PhysRevA.97.042347}. In an equivalent entanglement distillation version, Alice prepares an entangled state of the form
\begin{equation}
{\left| \phi  \right\rangle _{AB}} = \frac{1}{{\sqrt 2 }}\left( {{{\left| 0 \right\rangle }_A}{{\left| 0 \right\rangle }_B} + {{\left| 1 \right\rangle }_A}{{\left| 1 \right\rangle }_B}} \right),
\label{ea1}
\end{equation}
where she chooses one of photons to measure either in $\left\{ {\sigma _X^A,\sigma _Y^A,\sigma _Z^A} \right\}$, the other one is then sent to Bob, who chooses to measure in $\left\{ {\sigma _X^B,\sigma _Y^B,\sigma _Z^B} \right\}$. According to Eq.~(\ref{e2}), the optimization problem is
\begin{equation}
{\mathop{\rm minimize}\nolimits}: {C_L} = \sum\limits_{\kappa ,\zeta } {{\mathop{\rm Tr}\nolimits} {{\left( {\sigma _\kappa ^A \otimes \sigma _\zeta ^B{\rho _{AB}}} \right)}^2}},
\label{ea2}
\end{equation}
where ${\forall \left\{ {\kappa ,\zeta } \right\} \in \left\{ {X,Y} \right\}}$. We are allowing $\rho _{AB}$ to be arbitrary, which implies that Eve can perform any arbitrary operations on the states transmitted between Alice and Bob, and hence the bound is valid for any collective attacks.

Based on Eq.~(\ref{e1}), the error operator in the $Z$ basis is given by
\begin{equation}
{\hat e_{ZZ}} = \frac{1}{2}\left( {\openone \otimes \openone - \sigma _Z^A \otimes \sigma _Z^B} \right).
\label{ea3}
\end{equation}
Except for the case when Alice use the $Y$ basis to prepare her states, the probabilities $P_{{\alpha _i},{\chi _j}}$ can be simulated by
\begin{equation}
{P_{{\alpha _i},{\chi _j}}} = \frac{1}{2}\left[ {\left( {1 - {e_\alpha }} \right){{\left| {\left\langle {\alpha _i^A} \right|\left. {\chi _j^B} \right\rangle } \right|}^2} + {e_\alpha }{{\left| {\left\langle {\alpha _{i \oplus 1}^A} \right|\left. {\chi _j^B} \right\rangle } \right|}^2}} \right],
\label{ea4}
\end{equation}
where $\alpha  \in \left\{ {X,Z} \right\},\chi  \in \left\{ {X,Y,Z} \right\}$, and the symbol $ \oplus $ denotes the modulo two addition. The factor 1/2 is the conditional probability that the state $\left| {{\alpha _i^A}} \right\rangle$ is sent, given that it is prepared in $\alpha$ basis. $e_\alpha$ is the bit-flip rate of state  $\left| {{\alpha _i^A}} \right\rangle$, which is caused by the channel noise along the transmission. In Eq.~(\ref{ea4}), the first (second) term models the probability that Bob detects a state $\left| {\chi _j^B} \right\rangle $ after he received a state $\left| {\alpha _i^A} \right\rangle $ ($\left| {\alpha _{i \oplus 1}^A} \right\rangle $) sent from  Alice.

Due to $\left\langle {{\sigma _Y} \otimes {\sigma _Y}} \right\rangle  =  - 1$ for the state ${\left| \phi  \right\rangle _{AB}}$, Alice actually sends Bob a state $\left| {{Y_{i \oplus 1}}} \right\rangle$ after she got a state $\left| {{Y_{i}}} \right\rangle$ in the equivalent entanglement-based protocol. Thus, the the probabilities $P_{{Y _i},{\chi _j}}$, when Alice sends states in $Y$ basis, should be given by
\begin{equation}
{P_{{Y_i},{\chi _j}}} = \frac{1}{2}\left[ {{e_y}{{\left| {\left\langle {Y_i^A} \right|\left. {\chi _j^B} \right\rangle } \right|}^2} + \left( {1 - {e_y}} \right){{\left| {\left\langle {Y_{i \oplus 1}^A} \right|\left. {\chi _j^B} \right\rangle } \right|}^2}} \right].
\label{ea5}
\end{equation}

The QBERs ${e_{ZZ}}$, ${e_{XX}}$, ${e_{XY}}$, ${e_{YX}}$, and ${e_{YY}}$ can be extracted from the experiment, but not all of them equal to the corresponding bit-flip error rate as $\beta  \ne 0$. Here, we assume that they are all less than 0.5 (if not, Bob can simply flip his bits). For the convenience of experiment and simulation, we list the probabilities $P_{{\alpha _i},{\chi _j}}$ in detail as follows:
\begin{eqnarray}
\begin{aligned}
{P_{{Z_i},{\kappa _j}}} &= \frac{1}{4}; {P_{{\kappa _i},{Z_j}}} = \frac{1}{4};\\
{P_{{X_i},{X_i}}} &= \frac{1}{4}\left[ {1 + \cos \beta \left( {1 - 2{e_x}} \right)} \right]=\frac{1}{2}\left( {1 - {e_{XX}}} \right);\\
{P_{{X_i},{Y_i}}} &= \frac{1}{4}\left[ {1 - \sin \beta \left( {1 - 2{e_x}} \right)} \right]= \frac{1}{2}{e_{XY}};\\
{P_{{Y_i},{Y_i}}} &= \frac{1}{4}\left[ {1 - \cos \beta \left( {1 - 2{e_y}} \right)} \right]= \frac{1}{2}{e_{YY}};\\
{P_{{Y_i},{X_i}}} &= \frac{1}{4}\left[ {1 - \sin \beta \left( {1 - 2{e_y}} \right)} \right]= \frac{1}{2}{e_{YX}};\\
{P_{{X_i},{X_{j \ne i}}}} &= \frac{1}{4}\left[ {1 - \cos \beta \left( {1 - 2{e_x}} \right)} \right]=\frac{1}{2}{e_{XY}};\\
{P_{{X_i},{Y_{j\ne i}}}} &= \frac{1}{4}\left[ {1 + \sin \beta \left( {1 - 2{e_x}} \right)} \right]= \frac{1}{2}\left( {1 - {e_{XX}}} \right);\\
{P_{{Y_i},{Y_{j\ne i}}}} &= \frac{1}{4}\left[ {1 + \cos \beta \left( {1 - 2{e_y}} \right)} \right]= \frac{1}{2}\left( {1 -{e_{YY}}} \right);\\
{P_{{Y_i},{X_{j\ne i}}}} &= \frac{1}{4}\left[ {1 + \sin \beta \left( {1 - 2{e_y}} \right)} \right]= \frac{1}{2}\left( {1 -{e_{YX}}} \right).
\label{ea6}
\end{aligned}
\end{eqnarray}

\begin{figure}[t]
\centering
\includegraphics[width=0.48\textwidth]{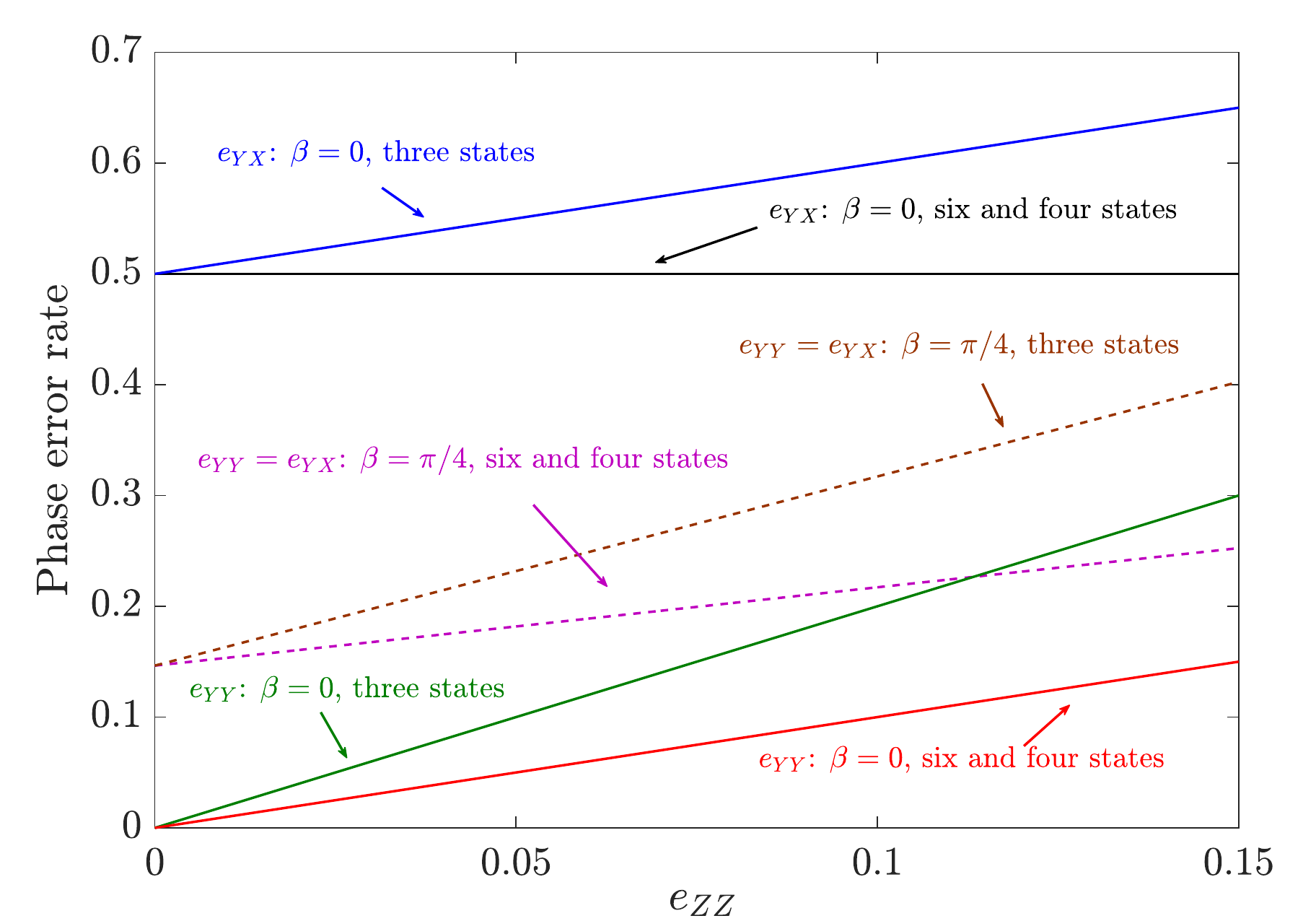}
\caption{\label{figa1}(Color online) The upper bound on the phase error rates $e_{YY}$ and $e_{YX}$ plotted as a function of the quantum error rates $e_{ZZ}$ when Alice sends six, four, and three states to Bob.}
\end{figure}

When the RFI QKD protocol with three states is applied, the fictious bit error rates $e_{YX}$ and $e_{YY}$ can also be well bounded using the follows optimization problem:
\begin{eqnarray}
\begin{aligned}
{\mathop{\rm maximize}\nolimits} :{e_{YX}} &= {\mathop{\rm Tr}\nolimits} \left( {{{\hat E}_{YX}}{\rho _{AB}}} \right);\\
{\mathop{\rm maximize}\nolimits} :{e_{YY}} &= {\mathop{\rm Tr}\nolimits} \left( {{{\hat E}_{YY}}{\rho _{AB}}} \right),
\label{ea7}
\end{aligned}
\end{eqnarray}
where
\begin{eqnarray}
\begin{aligned}
{{\hat E}_{YX}} = \frac{1}{2}\left( {\openone \otimes \openone - \sigma _Y^A \otimes \sigma _X^B} \right),\\
{{\hat E}_{YY}} = \frac{1}{2}\left( {\openone \otimes \openone - \sigma _Y^A \otimes \sigma _Y^B} \right).
\label{ea8}
\end{aligned}
\end{eqnarray}
In Fig.~\ref{figa1}, we show the dependence of $e_{YX}$ and $e_{YY}$ on $e_{ZZ}$ when Alice transmits different states at $\beta=0$ (solid lines) and $\beta=\pi/4$ (dashed lines). In simulations, it is noted that we treat the bit-flip rates in different bases equivalently for simplicity, i.e., ${e_z} = {e_x} = {e_y} = {e_{ZZ}}$. It is evident that for sending six and four states at $\beta=0$, $e_{ZZ}=e_{YY}$ and $e_{YX}=0.5$, as respectively indicated by the red solid line and black solid line. This is expected for RFI QKD protocol when no misalignment of reference frame occurs. However, the phase error rate $e_{YX}$ and $e_{YY}$ increase faster when Alice sends three states, resulting in lower $C_L$ as shown in Fig.~\ref{fig1}(a) and lower error tolerance as shown in Fig.~\ref{fig2}(b).

\section{Channel model}

\subsection{For the single-photon source}
We consider the channel model proposed by Ref.~\cite{PhysRevA.90.052314,PhysRevA.92.042319}, where the conditional probability that Bob obtain $j$ when he chooses $\chi $ basis for measurement given that Alice sent him the state $\left| {{\alpha _i}} \right\rangle$ can be written as
\begin{eqnarray}
\begin{aligned}
{V_{\left. {{\chi _j}} \right|{\alpha _i}}} = &\eta {T_{\left. {{\chi _j}} \right|{\alpha _i}}}\left( {1 - {e_d}} \right) + \left( {1 - \eta } \right){e_d}\left( {1 - {e_d}} \right)\\
& + \frac{1}{2}\left[ {\eta {e_d} + \left( {1 - \eta } \right)e_d^2} \right],
\label{ea9}
\end{aligned}
\end{eqnarray}
where $e_d$ is the dark count rate of SPD and $\eta$ denotes the total transmittance of the system. The term ${T_{\left. {{\chi _j}} \right|{\alpha _i}}}$ is the theoretical probability that Bob measures the state $\left| {{\alpha _i}} \right\rangle$ and obtains the bit value $j$ when he chooses $\chi$ basis. It is calculated by
\begin{eqnarray}
{T_{\left. {{\chi _j}} \right|{\alpha _i}}} = {\left| {\left\langle {{\alpha _i}} \right|\left. {{\chi _j}} \right\rangle } \right|^2}.
\label{ea10}
\end{eqnarray}
Then the single-photon gain and the QBER are given by, respectively,
\begin{eqnarray}
\begin{aligned}
Q_{\alpha \chi }^1 &= \frac{1}{2}\left( {{V_{\left. {{\chi _0}} \right|{\alpha _0}}} + {V_{\left. {{\chi _1}} \right|{\alpha _1}}} + {V_{\left. {{\chi _1}} \right|{\alpha _0}}} + {V_{\left. {{\chi _0}} \right|{\alpha _1}}}} \right),\\
E_{\alpha \chi }^1 &= \min \left[ {\tilde E_{\alpha \chi }^1,1 - \tilde E_{\alpha \chi }^1} \right],
\label{ea11}
\end{aligned}
\end{eqnarray}
where
\begin{eqnarray}
\begin{aligned}
\tilde E_{\alpha \chi }^1 &= {e_o}\left( {1 - 2e_{\alpha \chi }^1} \right) + e_{\alpha \chi }^1,\\
e_{\alpha \chi }^1 & = \frac{{{V_{\left. {{\chi _1}} \right|{\alpha _0}}} + {V_{\left. {{\chi _0}} \right|{\alpha _1}}}}}{{2Q_{\alpha \chi }^1}}.
\label{ea12}
\end{aligned}
\end{eqnarray}
The factor 1/2 in Eq.~(\ref{ea11}) denotes the probability of Alice preparing quantum states ${\left| {{\alpha _0}} \right\rangle }$ or ${\left| {{\alpha _1}} \right\rangle }$, and $e_o$ in Eq.~(\ref{ea12}) is the optical intrinsic error rate \cite{PhysRevA.86.052305,PhysRevA.89.052333}. For simplicity, we assume $E_{\alpha \chi }^1 \le 0.5$ in Eq.~(\ref{ea11}), if not, either Alice or Bob flips her or his bit strings to make it hold.

\subsection{For the WCS with decoy technique}
In this case, according to the decoy state method \cite{PhysRevA.92.042319}, the overall gain, given that Alice sends a state $\left| {{\alpha _i}} \right\rangle$ using $k \in \mathcal{K}$ intensity and Bob obtains a state $\left| {{\chi _j}} \right\rangle$, can be written as
\begin{eqnarray}
\begin{aligned}
{Q_{k,{\alpha _i},{\chi _j}}} &= \sum\limits_{n = 0}^\infty  {{Y_n}\frac{{{k^n}}}{{n!}}} {e^{ - k}}\\
&{\rm{ = }}\frac{{\rm{1}}}{{\rm{2}}}\left\{ {{\rm{1 + }}D\left[ {{e^{\left( { - \eta  + a} \right)k}} - {e^{ - ak}} - D{e^{ - \eta k}}} \right]} \right\},
\label{ea13}
\end{aligned}
\end{eqnarray}
where we use the notations
\begin{equation}
a = \eta {T_{\left. {{\chi _j}} \right|{\alpha _i}}},D = 1 - {e_d}.
\label{ea14}
\end{equation}
According to the above formula, we can obtain the overall gain, the overall error gain, and the averge of observed error rate in the $\alpha$ basis. They are given by, respectively
\begin{equation}
\begin{aligned}
Q_{\alpha \chi }^k &= \frac{1}{2}\left( {{Q_{k,{\alpha _0},{\chi _0}}} + {Q_{k,{\alpha _1},{\chi _0}}} + {Q_{k,{\alpha _1},{\chi _1}}} + {Q_{k,{\alpha _0},{\chi _1}}}} \right),\\
W_{\alpha \chi }^k &= {e_o}\left( {Q_{\alpha \chi }^k - 2\tilde W_{\alpha \chi }^k} \right) + \tilde W_{\alpha \chi }^k,\\
E_{\alpha \chi }^k &= \min \left[ {\tilde E_{\alpha \chi }^k,1 - \tilde E_{\alpha \chi }^k} \right],
\label{ea15}
\end{aligned}
\end{equation}
where
\begin{equation}
\begin{aligned}
\tilde W_{\alpha \chi }^k &= \frac{1}{2}\left( {{Q_{k,{\alpha _1},{\chi _0}}} + {Q_{k,{\alpha _0},{\chi _1}}}} \right),\\
\tilde E_{\alpha \chi }^k& = {{W_{\alpha \chi }^k} \mathord{\left/
 {\vphantom {{W_{\alpha \chi }^k} {Q_{\alpha \chi }^k}}} \right.
 \kern-\nulldelimiterspace} {Q_{\alpha \chi }^k}}.
\label{ea16}
\end{aligned}
\end{equation}

Considering the probabilities that Alice prepares her state in the $\alpha$ basis, $\Pr _\alpha ^A$, its mean photon number $k$, $p_k$, and Bob measures this state in the $\chi$ basis, $\Pr _\chi ^B$, we can calculated the number of detected pulses, ${n_{\alpha \chi ,k}}$ and the number of bit errors, ${m_{\alpha \chi ,k}}$, when Alice prepares her state in the $\alpha$ basis with intensity $k$ and Bob measures it in the $\chi$ basis. They are given by
\begin{equation}
\begin{aligned}
{n_{\alpha \chi ,k}} &= N{p_k}\Pr _\alpha ^A\Pr _\chi ^BQ_{\alpha \chi }^k,\\
{m_{\alpha \chi ,k}} &= N{p_k}\Pr _\alpha ^A\Pr _\chi ^BW_{\alpha \chi }^k.
\label{ea17}
\end{aligned}
\end{equation}
Thus, the overall number of detected pulses and the overall number of bit errors for all intensity levels are 
\begin{eqnarray}
\begin{aligned}
{n_{\alpha \chi }} &= \sum\nolimits_{k \in K} {{n_{\alpha \chi ,k}}} ,\\
{m_{\alpha \chi }} &= \sum\nolimits_{k \in K} {{m_{\alpha \chi ,k}}} .
\label{ea19}
\end{aligned}
\end{eqnarray}

\nocite{*}

\bibliography{apssamp}

\end{document}